\documentclass[lettersize,journal]{IEEEtran}
\usepackage{amsfonts}
\usepackage{algorithmic}
\usepackage{algorithm}
\usepackage{array}
\usepackage[caption=false,font=normalsize,labelfont=sf,textfont=sf]{subfig}
\usepackage{textcomp}
\usepackage{stfloats}
\usepackage{url}
\usepackage{verbatim}
\usepackage{graphicx}
\usepackage{xcolor}
\usepackage{cite}
\usepackage{amsmath}
\usepackage{gensymb}
\usepackage{graphicx}
\usepackage{float}
\usepackage{txfonts}
\usepackage{authblk}

\begin{document}

\title{Improved Modeling of Quasi-Static Thermal and Optical Response of Lumped-Element Aluminum Manganese KIDs}

\author[1]{Adriana Gavidia}
\author[1]{Sunil Golwala}
\author[2]{Andrew D. Beyer}
\author[2]{Daniel Cunnane}
\author[2]{Peter K. Day}
\author[2]{Fabien Defrance}
\author[2]{Clifford F. Frez}
\author[3, 4]{Xiaolan Huang}
\author[1, 6]{Junhan Kim}
\author[1]{Jean-Marc Martin}
\author[1]{Jack Sayers}
\author[3]{Shibo Shu}
\author[1, 5, 7]{Shiling Yu}
\author[1]{Yann Sadou}

\affil[1]{\footnotesize Division of Physics, Mathematics, and Astronomy, California Institute of Technology, Pasadena, CA, USA, 91125}
\affil[2]{Jet Propulsion Laboratory, California Institute of Technology, 4800 Oak Grove Ave., Pasadena, CA, USA, 91109}
\affil[3]{Institute of High Energy Physics, Chinese Academy of Sciences, Beijing, China, 100049}
\affil[4]{Shanghai Normal University, Shanghai, China, 201418}
\affil[5]{National Astronomical Observatories of China, Beijing, China, 100101}
\affil[6]{Korea Advanced Institute for Science and Technology, Daejeon, 34141, Republic of Korea}
\affil[7]{University of the Chinese Academy of Sciences, Beijing, China, 101408}

\markboth{Journal of \LaTeX\ Class Files,~Vol.~14, No.~8, August~2021}%
{Shell \MakeLowercase{\textit{et al.}}: A Sample Article Using IEEEtran.cls for IEEE Journals}

\maketitle

\begin{abstract}
We report on the optical characterization of the AlMn kinetic inductance detectors (KIDs) in development for use in the Next-generation Extended Wavelength-MUltiband Sub/millimeter Inductance Camera (NEW-MUSIC)~\cite{Golwala2024} on the Leighton Chajnantor Telescope (LCT).  NEW-MUSIC will cover 80-420 GHz, split into six spectral bands, with polarimetry. This broad spectral coverage will enable study of a range of scientific topics such as the accretion and feedback in galaxies and galaxy cluster evolution via the Sunyaev-Zeldovich effect, the transient synchrotron emission from the explosive deaths of massive stars and other time-domain phenomena, and dusty sources from low to high redshift (with polarization).  Al KIDs have already been demonstrated for bands 2-5 \cite{Golwala2024}. AlMn KIDs will be used for the 90~GHz band, as Al's pair-breaking energy is too high. However, AlMn has only barely been explored as a KID material.  To this end, we first improved the modeling techniques used for Al KIDs within BCS theory by eliminating the use of analytical approximations for the expressions of the complex conductivity and found  these changes reduced fit parameter degeneracy in the analysis of AlMn. Then, we tested the addition of a gap smearing parameter, a standard extension to BCS theory in use for high kinetic inductance materials, and found it did not improve the fits.  
 
\end{abstract}


\section{Introduction}
\IEEEPARstart{T}{he} study of a variety of astrophysical phenomena will be vastly expanded by broadband trans-mm observations from future large-aperture (30-50 m) ground-based telescopes (CSST, CMB-HB, AtLAST) in the 80-420 GHz range. This spectral range includes extremely energetic phenomena where transient and time-domain synchrotron emission shows peaks and spectral breaks from events at multiple distance scales, like core-collapse supernovae, gamma-ray bursts, tidal disruption events, and active galactic nuclei (AGN). Additionally, the Sunyaev-Zeldovich (SZ) effect, bright in this spectral range,  can probe the total thermal content of galaxy clusters and the circumgalactic medium of field galaxies, providing critical insight into their evolutionary history. Multi-band data through the trans-mm regime is necessary to isolate the thermal and kinetic effects, relativistic corrections, and contaminating foreground and background sources. The Next-generation Extended Wavelength-MUltiband Sub/millimeter Inductance Camera (NEW-MUSIC)~\cite{Golwala2024} on the 10.4~m Leighton Chajnantor Telescope (LCT) will prototype and pathfind for these types of observations by providing 2.4 octaves of co-pointed spectropolarimetric coverage in six spectral bands 80-420 GHz (0.7-3.8 mm).

To achieve this broad frequency coverage, light is received through hierarchical, phased arrays of polarization-sensitive slot-dipole antennas, which incorporate low-loss, hydrogenated amorphous silicon (a-Si:H) dielectric in the microstripline feed network.  The light is summed in a frequency-selective manner and split into six spectral bands over 80-420 GHz by photolithographic in-line bandpass and lowpass filters~\cite{newmusic_ltd19_shu,newmusic_ltd20_martin,newmusic_ltd21_huang}.  The spectrally filtered light is routed to aluminum (Al) or aluminum-manganese (AlMn) microstrip-coupled, parallel-plate capacitor, lumped-element KIDs (MS-PPC-LEKIDs). We demonstrated previously that our Al KID design is limited by fundamental noise (photon noise and generation-recombination (GR) noise) at $>100$~Hz for a wide range of loadings from 130 to 370 GHz \cite{Golwala2024}, and we have recently shown this performance extends down to 0.1-1~Hz~\cite{newmusic_ltd21_hempel-costello}.  However, Al, especially in thin films, has $2\Delta = 3.76\,k_B\,T_c$, too high to be suitable for band 1, whose lower edge is at $\approx$80~GHz. AlMn is a promising candidate for this band, as it can be fabricated with nearly identical processing as pure Al and allows for control of the $T_c$ with varying levels of Mn doping \cite{Young2002} \cite{Young2004} \cite{Deiker2004}. 

The modeling of AlMn devices may be more complex than Al, as films of this material have been shown to have a slight broadening of the superconducting gap \cite{Kaiser1970} \cite{ONeil2008} \cite{ONeil2010}, which is not accounted for within the BCS Mattis-Bardeen model we employ. However, the standard BCS Mattis-Bardeen model has been successfully applied to AlMn devices \cite{Jones2017}. Before exploring models outside of BCS theory, we therefore first improve upon the modeling of the shift in frequency due to changes in quasi-particle density used for the Al devices in \cite{Golwala2024}, which employed analytic approximations to obtain the complex conductivity within Mattis-Bardeen theory. Instead, we implement fully numerical models. We then assess whether the addition of a parameter to account for any broadening of the superconducting gap is necessary within our fits.

\section{Mattis-Bardeen Fits to AlMn}  
\subsection{Mattis-Bardeen Theory}
We use Mattis-Bardeen theory \cite{MB1958} to determine the 
relationship between the quasi-particle density ($n_{\text{qp}}$) of a thin superconducting film and its surface impedance. This is a crucial step in characterizing the complex transmission through our system, which is directly measured by our readout electronics. We introduce a complex conductivity $\sigma = \sigma_1 - j\sigma_2$ to describe the superconducting state \cite{Glover1957}. From Mattis-Bardeen theory, the following integrals hold: 
\begin{equation}
    \sigma_1 = \frac{2 \sigma_n}{\hbar \omega} \int_\Delta^\infty \frac{[f(E) - f(E+\hbar \omega)](E^2 + \Delta^2 + \hbar \omega E)}{\sqrt{(E^2 - \Delta^2) \left[(E + \hbar \omega)^2 - \Delta^2 \right]}} dE
    \label{eq:sigma1}
\end{equation}
\begin{equation}
    \sigma_2 = \frac{\sigma_n}{\hbar \omega} \int_{\Delta - \hbar \omega}^{\Delta} \frac{[1 - 2f(E + \hbar \omega](E^2 + \Delta^2 + \hbar \omega E)}{\sqrt{(\Delta^2 - E^2) - \Delta^2}} dE
    \label{eq:sigma2}
\end{equation}
for $\hbar \omega < \Delta$, where $\omega = 2\pi f$ is the angular frequency, $\Delta$ is the gap energy, $\sigma_n$ is the normal state conductivity, and $f(E)$ is the distribution function for quasi-particles. In this case, under the assumption that quasi-particles are in thermal equilibrium, $f(E)$ is given by the Fermi-Dirac distribution, 
\begin{equation}
    f(E) = \frac{1}{1 + e^\frac{E-\mu^\star}{k_B T}}.
    \label{eq:FD}
\end{equation}
where $\mu^\star$ is the chemical potential and $k_B$ is the Boltzmann constant.
The corresponding quasi-particle density is given by 
\begin{equation}
    n_{\text{qp}} = 4N_0 \int_\Delta^\infty dE \frac{E}{\sqrt{E^2 - \Delta^2}} f(E),
    \label{eq:nqp}
\end{equation}
where $N_0$ is the single-spin density of electron states at the Fermi energy. 

The fractional change in the complex conductivity in response to changes in $n_{\text{qp}}$ is what is used to predict detector response, so we define this fractional change with respect to its value with no optical loading, which we will refer to as dark, and at zero temperature. From Eqs. \ref{eq:FD} and \ref{eq:nqp}, we see that $n_{\text{qp}}$ vanishes exponentially as $T \rightarrow 0$ ($\mu^\star = 0$ when there is no optical loading). The real component of the complex conductivity, $\sigma_1$, also follows this behavior, so $\sigma_1^{\text{dark}}(0) = 0$. The imaginary component of the complex conductivity, $\sigma_2$, approaches a constant value of $\sigma_2^{\text{dark}}(0) = (\pi \Delta_0 / \hbar \omega) \sigma_n$ due to the kinetic inductance of Cooper pairs. So, we have 
\begin{equation}
    \sigma_0 = \sigma_1^{\text{dark}}(0) - j \sigma_2^{\text{dark}}(0) = -j \frac{\pi \Delta_0}{\hbar \omega} \sigma_n
\end{equation}
and the fractional change in the complex conductivity is 
\begin{equation}
    \frac{\delta \sigma}{|\sigma|} \equiv \frac{\sigma -\sigma_0}{|\sigma_0|} = (\kappa_1 - j\kappa_2)n_{\text{qp}}
\end{equation}
where we defined $\kappa_1 \equiv \sigma_1/(\sigma_0 n_{\text{qp}})$ and $\kappa_2 \equiv \sigma_2/(\sigma_0 n_{\text{qp}})$. The fractional frequency shift is then a function of $\kappa_2$ and the $n_{\text{qp}}$: 
\begin{equation}
    \frac{\delta f_{\text{res}}}{f_0} = -\frac{\alpha}{2} \kappa_2 n_{\text{qp}}.
    \label{eq:fres_shift}
\end{equation}
Under optical loading, the quasi-particle density is given by
\begin{multline}
    n_{\text{qp}} = \bigg[\frac{\eta_\text{opt} \eta_{\text{pb}} k_B \Delta\nu (T_{\text{load}} + T_{\text{exc}})}{RV \Delta_0} + n_{\text{qp,th}}^2 \\ 
    + \frac{1}{R \tau_\text{max}} \left(n_{\text{qp,th}} + \frac{1}{4R \tau_{\text{max}}}\right) \bigg]^{1/2} - \frac{1}{2R \tau_{\text{max}}},
    \label{eq:nqp_opt}
\end{multline}
where $\delta f_{\text{res}} = f_{\text{res}}(T_{\text{bath}}, T_{\text{load}} + T_\text{exc}) - f_0$, $f_0 = f_\text{res}(T_\text{bath} = 0, T_\text{load} + T_\text{exc} = 0)$, $T_\text{load}$ is the temperature of the blackbody load placed in front of the vacuum window, $T_\text{bath}$ is the temperature at the detector stage, $T_\text{exc}$ accounts for a fixed additional excess optical loading originating from the cryostat, $n_\text{qp,th}$ is the thermal quasi-particle density, $\alpha$ is the kinetic inductance fraction, $\eta_\text{opt}$ is the total optical efficiency, $\eta_\text{pb}$ is the pair-breaking efficiency, $\Delta \nu$ is the detector bandwidth, $V=3224$ $\mu \text{m}^3$ is the KID inductor volume, $n_\text{th}$ is the thermal quasi-particle density, $\tau_\text{max} = 400$ $\mu \text{s}$ is the maximum quasi-particle life time  \cite{Kozorezov2008, Barends2008, Barends2009}, $R = 2\Delta_0^2 / [N_0 \tau_0 (k_B T_c)^3]$ is the recombination rate per unit density of quasi-particles, $T_c$ is the critical temperature of the KID inductor material, and $\tau_0$ = 438 ns is the characteristic electron-phonon interaction time. 

In the dark scenario, Eq.~\ref{eq:fres_shift} reduces to 
\begin{equation}
    \frac{\delta f_{\text{res}}}{f_0} = -\frac{\alpha}{2} \kappa_2 n_{\text{qp,th}}.
    \label{eq:fres_dark}
\end{equation}
To calculate the thermally generated quasi-particles, we use Eq.~\ref{eq:nqp}, taking $\mu^\star = 0$ in the Fermi-Dirac distribution (Eq.~\ref{eq:FD}), thus isolating the thermal contribution, and we assume $\Delta = \Delta_0$. 

Analytic forms of Eqs.~\ref{eq:sigma1}, \ref{eq:sigma2}, and \ref{eq:nqp} can be obtained in the limit that $k_BT \ll \Delta_0$, $\hbar\omega \ll \Delta$ and $\exp[-(E-\mu^\star)/k_B T] \ll 1$. These analytic approximations were employed in a previous analysis of our AlMn devices and were unable to accurately model the measured $\delta f_\text{res}/f_0$ data \cite{Golwala2024}. Because we seek to assess how well our thin films of AlMn follow BCS theory, we first improve our Mattis-Bardeen modeling by calculating $\sigma_1$, $\sigma_2$, and $n_\text{qp}$ fully numerically, rather than using their analytic approximations, before considering any models outside of BCS theory. 

\begin{figure}
    \centering
    \includegraphics[width=0.99\linewidth]{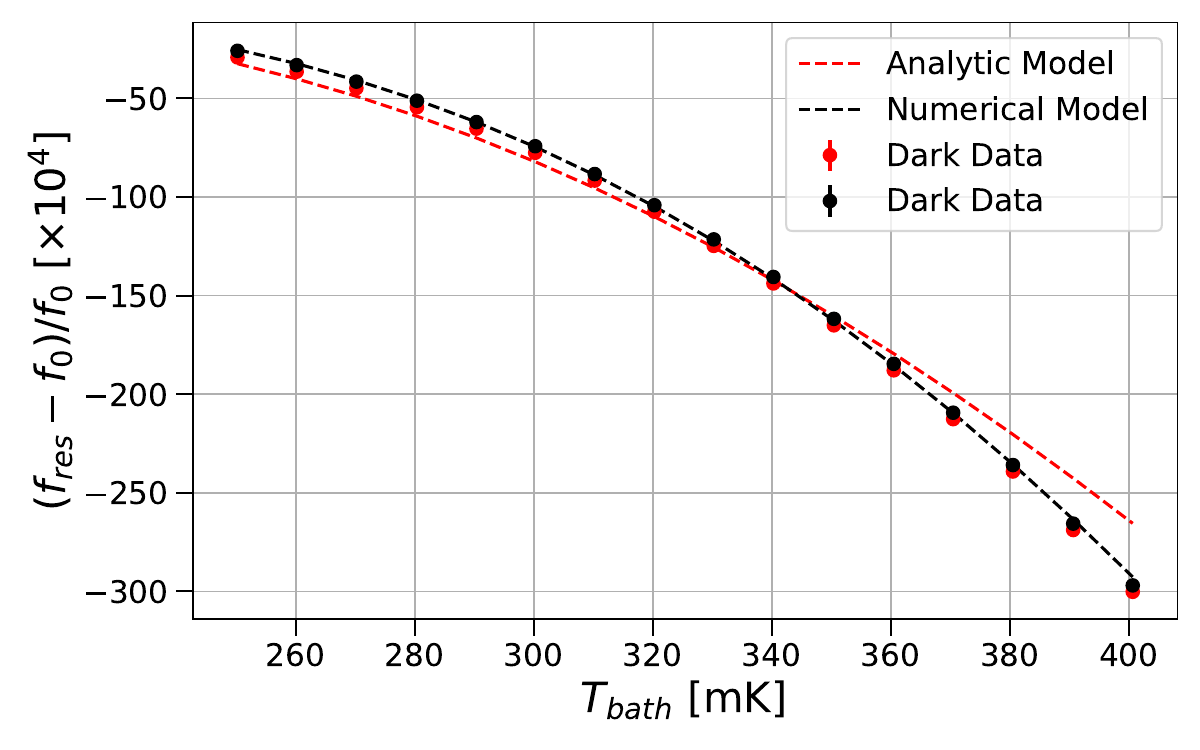}
    \caption{Measured $\delta f_\text{res}/f_0$ data points (black and red dotted curves) taken with no optical loading for a single AlMn KID resonator. The fitted model to the data using a fully numerical implementation of Mattis-Bardeen theory is shown as the dashed black line and the analytical implementation is shown as the dashed red line. }
    \label{fig:MBfit_dark}
\end{figure}

\subsection{Dark Scenario} \label{sec:dark}
To calibrate the frequency response to changes in $n_{\text{qp,th}}$, we use a vector network analyzer (VNA) to measure $S_{21}(f)$ over the frequency range containing the resonances at a range of $T_{\text{bath}}$ values, with no optical loading (dark). We fit these resonance scans using standard techniques (e.g. \cite{Gao2008}). This provides the fractional frequency shift of each resonator as a function of bath temperature. Then, this processed data is fit to Eq.~\ref{eq:fres_dark} using a least squares minimization routine, performing the complex conductivity integrals (Eqs.~\ref{eq:sigma1}, \ref{eq:sigma2}, \ref{eq:nqp}) fully numerically. This provides constraints on $\alpha$ and $\Delta_0$ (assuming $\Delta = \Delta_0$). 

\begin{figure}
    \centering
    \includegraphics[width=0.99\linewidth]{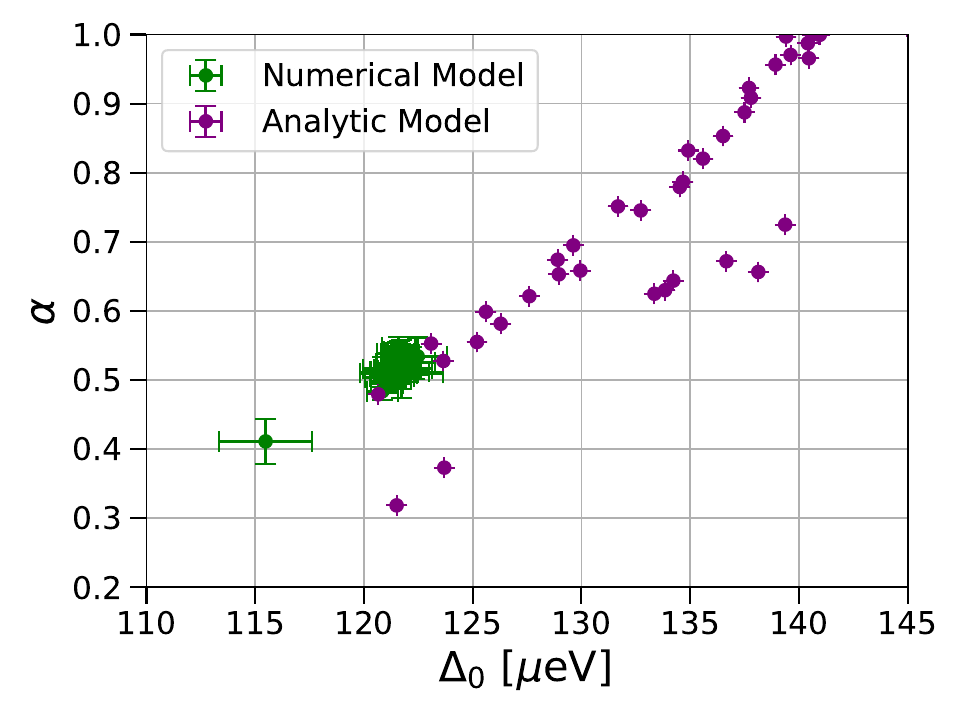}
    \caption{Recovered values for $\alpha$ and $\Delta_0$ for all AlMn resonators from fits employing analytic approximations of Mattis-Bardeen theory (Eqs. \ref{eq:sigma1} and \ref{eq:sigma2}) in purple and from fully numerical fits in green. The degeneracy between these two parameters vanishes when the numerical technique is used.}
    \label{fig:alpha_delta}
\end{figure}

The fractional frequency shift model obtained by the fit for a single resonator is shown as the dashed black curve in Figure~\ref{fig:MBfit_dark} and is compared to measured data. We see excellent agreement between the fitted model and the measured data. To compare with the analysis method employing analytical approximations, we also plot the fitted model obtained from such a technique in red. Note that the red and black data points are identical $f_\text{res}$ measurements, but their corresponding $\delta f_\text{res}/f_0$ curves differ because they are scaled by the $f_0$ values obtained by their respective fitting techniques. We see the fit quality improves significantly in the numerical case, by a factor of $\chi_\text{analytical}^2/\chi_\text{numerical}^2 \sim 50$. Additionally, the fits employing analytical approximations have a strong degeneracy between the recovered $\alpha$ and $\Delta_0$ values for all the resonators, as shown by the purple data points in Figure~\ref{fig:alpha_delta}. Conversely, the recovered $\alpha$ and $\Delta_0$ parameters from the numerical fits, shown as green data points, display no such degeneracy. 

To further explore possible modeling techniques for AlMn, we included a gap broadening term in the fits to the dark data \cite{Kaiser1970} \cite{ONeil2008} \cite{ONeil2010}. This manifests as a transformation performed on the gap energy: $\Delta \rightarrow \Delta + j\Gamma$ applied in Eqs.~\ref{eq:sigma1}, \ref{eq:sigma2}, and \ref{eq:nqp}, where $\Gamma$ is an additional free parameter in the fit. Including this effect in our modeling degraded the $\chi^2$ of the numerical analysis by a factor of $\sim$6, so we did not include it in the results presented here. Thus, we conclude the numerical forms of the Mattis-Bardeen equations are sufficient to model $\delta f_{\text{res}} / f_0$ for AlMn.

\begin{figure*}
    \centering
    \includegraphics[width=0.49\linewidth]{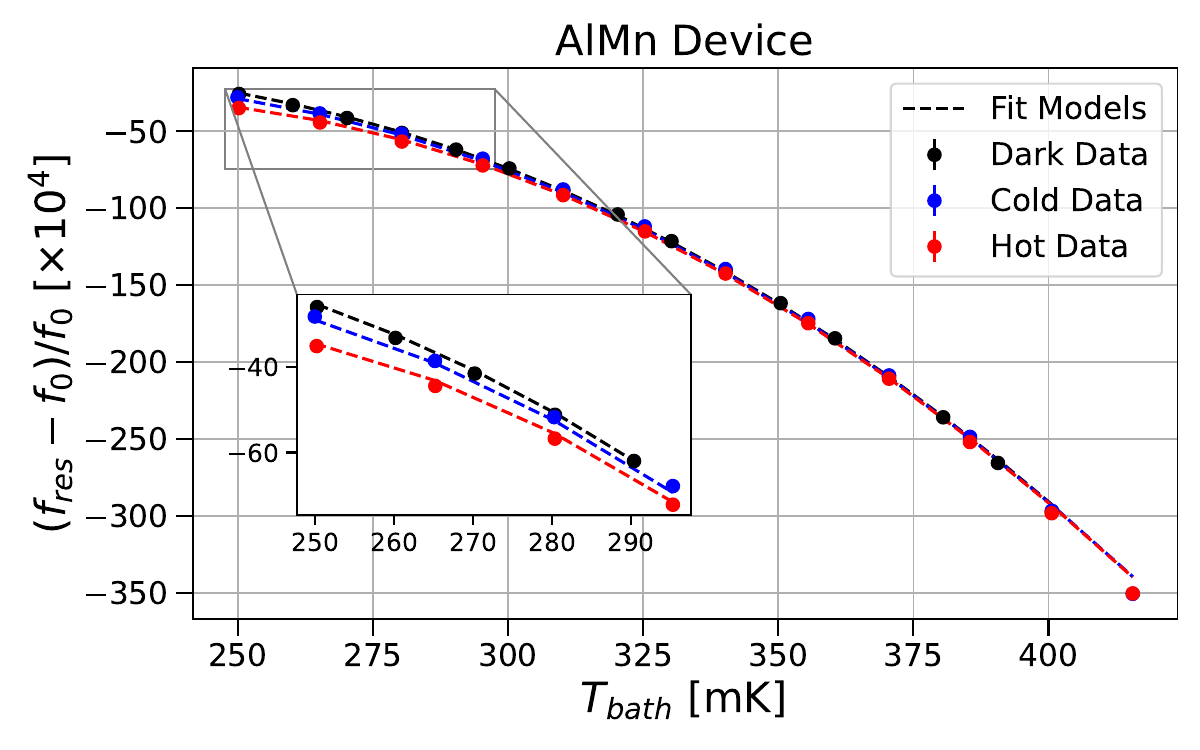}
    \includegraphics[width=0.49\linewidth]{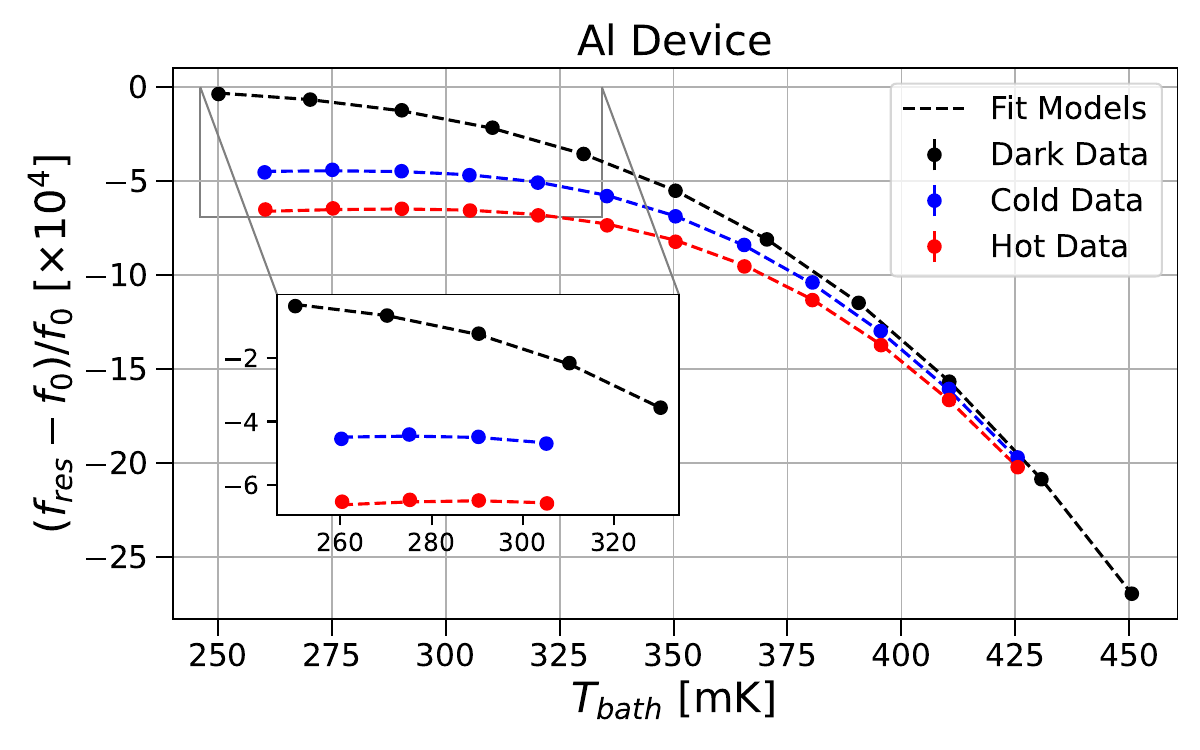}
    \caption{Typical $\delta f_\text{res}/f_0$ data sets used to fit for optical efficiency. Black, blue, and red data points and lines correspond to dark, cold, and hot data and fits. The plot on the left shows this for an AlMn resonator and on the right is for an Al resonator. The zoomed-in portion of the plots shows the response in the low temperature range where optically generated quasi-particles should dominate over thermally generated quasi-particles. The response level in this regime, in particular, the magnitude of the difference between the hot and cold curves, corresponds to the optical efficiency.}
    \label{fig:MBfit_optical}
\end{figure*}

\subsection{Optical Scenario}
We measured the optical efficiency of our AlMn devices by performing the same $S_{21}(f)$ scan, now alternating between a beam filling black body load cooled to the temperature of liquid nitrogen ($T_{\text{load}}$ = 77.36 K), denoted by ``cold'', and an ambient, room temperature black body load ($T_{\text{load}}$ = 294.15 K), denoted by ``hot'', in front of the cryostat window at each $T_\text{bath}$ value. The resulting $\delta f_{\text{res}}/f_0$ measurements are fit to the model given by Eq.~\ref{eq:fres_shift}, with the goal of obtaining the value of power incident on the KID from the microstripline, given by $P_{\text{opt}} = \eta_{\text{opt}} k_B T_{\text{load}} \Delta\nu$, for a single polarization. The Mattis-Bardeen fits to this data constrain $\eta_{\text{opt}}$ (Eq.~\ref{eq:nqp_opt}), however, we note that this parameter is completely degenerate with several other parameters: $\alpha$, $\Delta_0$, $\Delta\nu$, $\eta_{\text{pb}}$, $N_0$, $R$, and $V$. To deal with this, we supplement the hot/cold data with independent measurements of these other parameters. The values of $\alpha$ and $\Delta_0$ are set to those obtained by the fits to the dark data. We determine $\Delta\nu$ via Fourier Transform Spectroscopy (FTS). Then, given $\Delta \nu$ and $\Delta_0$, we obtain  $\eta_{\text{pb}}$ using the prescription outlined in \cite{Guruswamy2014}. The value of $N_0$ for aluminum that we assume is reported in the literature as $N_0 = 1.72 \times 10^{10}\ \mu\text{m}^{-3} \text{eV}^{-1}$ \cite{Gao2008} \cite{Mazin2005} \cite{McMillan1968}. Then, the KID inductor volume is taken to be 3224~$\mu \text{m}^3$ by design. Previous attempts to perform this measurement of optical efficiency for Al KIDS yielded values consistent with the expected design values for bands 2-5 \cite{Golwala2024}. 

Our numerical fits to the data taken for the hot and cold loads for a single AlMn resonator, using the model given by Eq.~\ref{eq:fres_shift}, are shown on the left in Figure~\ref{fig:MBfit_optical}, along with the corresponding dark fit from the previous section. We see good agreement between the optical model and data in the low temperature regime. Although this model tends to overestimate the frequency shift at higher temperatures, in general, the hot and cold fits converge to dark fit in the high temperature range where thermally generated quasi-particles dominate. We highlight the low temperature range of the fits, where the three data sets deviate from one another due to differing degrees of contribution from optically generated quasi-particles. For comparison, we show the same fits performed on an Al resonator on the right in Figure~\ref{fig:MBfit_optical}. A key difference between these results in the low temperature regime is that the $\delta f_{\text{res}}/f_0$ curve for Al flattens out in this region, indicating that it is in the steady state where optically generated quasi-particles dominate over thermally generated quasi-particles. The difference between the quasi-particle density, which is related to the fractional frequency shift via Eq.~\ref{eq:fres_shift}, under hot and cold loads in this steady state tells us about the optial efficiency. In particular, larger offsets generally correspond to higher optical efficiency. We see then that our ability to constrain optical efficiency is limited by the response of the device to optical loading. The small response of the AlMn resonator (compared to that of Al) prevented the fit from obtaining meaningful constraints on $\eta_\text{opt}$. 

The primary factor we believe is responsible for the small response of AlMn is the temperature range we probe. The $T_c$ of AlMn is lower than that of Al. As shown in Figure~\ref{fig:MBfit_optical}, we measure $\delta f_{\text{res}}/f_0$ over approximately the same range for Al and AlMn. The AlMn device does not reach the steady state regime where optically generated quasi-particles dominate in this $T_\text{bath}$ range, which may imply, given the lower $T_c$ of AlMn, we have not probed to low enough $T_{\text{bath}}$ to enter the regime where this occurs for AlMn.

\section{Conclusion}
We have demonstrated that, relative to the analytic implementation, a numerical implementation of Mattis-Bardeen theory (Eqs.~\ref{eq:sigma1}, \ref{eq:sigma2}, and \ref{eq:nqp}) vastly improves the quality of fit to fractional frequency shift vs.\ temperature measurements of our AlMn resonators, now yielding satisfactory agreement. Additionally, we found that, when we included a gap broadening term in our model, the fit quality degraded, implying that our AlMn films do not suffer noticeable gap smearing. This numerical model was applied to optical data in order to constrain the optical efficiency of the devices. Although the model adequately described the data, we were unable to obtain useful constraints on the optical efficiency because the device response to optical load is small in the 250-350~mK regime.

\section*{Acknowledgments}
This work has been supported by the JPL Research and Technology Development Fund, the National Aeronautics and Space Administration under awards 80NSSC18K0385 and 80NSSC22K1556, the Department of Energy Office of High-Energy Physics Advanced Detector Research program under award DE-SC0018126, and the Wilf Foundation. The research was carried out in part at the Jet Propulsion Laboratory, California Institute of
Technology, under a contract with the National Aeronautics and Space Administration (80NM0018D0004).

\bibliographystyle{IEEEtran}
\bibliography{refs}

\vfill

\end{document}